\documentclass[aps,pra,longbibliography,preprint,superscriptaddress,showpacs]{revtex4-1}
\usepackage{graphicx}
\usepackage{dcolumn}
\usepackage{bm}
 \usepackage{tensor}
 \usepackage{physics}
 \usepackage{color}
\usepackage{times}
\usepackage{siunitx}
\begin{document}


\title{Hermite-Gaussian mode sorter}

\author{Yiyu Zhou}
\email{yzhou62@ur.rochester.edu}
\affiliation{The Institute of Optics, University of Rochester, Rochester, New York 14627, USA}
\author{Jiapeng Zhao}
\affiliation{The Institute of Optics, University of Rochester, Rochester, New York 14627, USA}
\author{Zhimin Shi}
\email{zhiminshi@usf.edu}
\affiliation{Department of Physics, University of South Florida, Tampa, Florida 33620, USA}
\author{Seyed Mohammad Hashemi Rafsanjani}
\affiliation{The Institute of Optics, University of Rochester, Rochester, New York 14627, USA}
\affiliation{Department of Physics, University of Miami, Coral Gables, Florida 33146, USA}
\author{Mohammad Mirhosseini}
\affiliation{The Institute of Optics, University of Rochester, Rochester, New York 14627, USA}
\affiliation{Thomas J. Watson, Sr., Laboratory of Applied Physics, California Institute of Technology, Pasadena, California 91125, USA}
\author{Ziyi Zhu}
\affiliation{Department of Physics, University of South Florida, Tampa, Florida 33620, USA}
\author{Alan E. Willner}
\affiliation{Department of Electrical Engineering, University of Southern California, Los Angeles, California 90089, USA}
\author{Robert W. Boyd}
\email{boyd@optics.rochester.edu}
\affiliation{The Institute of Optics, University of Rochester, Rochester, New York 14627, USA}
\affiliation{Department of Physics, University of Ottawa, Ottawa, Ontario K1N 6N5, Canada}

\begin{abstract}
The Hermite-Gaussian (HG) modes, sometimes also referred to as transverse electromagnetic modes in free space, form a complete and orthonormal basis that have been extensively used to describe optical fields. In addition, these modes have been shown to be helpful to enhance information capacity of optical communications as well as to achieve super-resolution imaging in microscopy. Here we propose and present the realization of an efficient, robust mode sorter that can sort a large number of HG modes based on the relation between HG modes and Laguerre-Gaussian (LG) modes. We experimentally demonstrate the sorting of 16 HG modes, and our method can be readily extended to a higher-dimensional state space in a straightforward manner. We expect that our demonstration will have direct applications in a variety of fields including fiber optics, classical and quantum communications, as well as super-resolution imaging.
\end{abstract}

\maketitle

The transverse modes of electromagnetic waves have long been used in fundamental studies of beam propagation \cite{svelto1998principles}. Spatial mode decomposition of optical fields can facilitate the understanding and analyzing of optical beam in free space \cite{paterson2005atmospheric, andrews2005laser}, graded-index multi-mode fibers \cite{renninger2013optical, flaes2018robustness}, and waveguides \cite{ortega2003adaptive}. Beyond their advantage for theoretical work, the transverse degrees of freedom of photons are recognized as information resources for both classical and quantum information technologies because of the unbounded Hilbert space spanned by these spatial mode basis sets \cite{wang2012terabit, mair2001entanglement, mirhosseini2015high, wang2018towards}. The Hermite-Gaussian (HG) modes are the propagation-invariant modes in parabolic-index multi-mode fibers \cite{flaes2018robustness,shemirani2009principal} and closely resemble the communication modes of square apertures for free-space propagation \cite{rodenburg2015communicating}, which suggests the potential of HG modes in optical communications. In addition, very recently it has been shown that the detection of HG modes can beat "Rayleigh's curse" and realize super-resolution imaging \cite{tsang2016quantum}. It is proposed that spatial mode decomposition in the HG basis can reach the Cram\'er-Rao bound for resolving two closely located point sources, while the classical Fisher information of traditional imaging in the position basis inherently drops to zero. There have been various experimental investigations that propose alternative detection strategies to achieve a nonzero Fisher information for a small separation \cite{tham2017beating, paur2016achieving, yang2016far}. However, because of the lack of an efficient HG mode sorter, the Cram\'er-Rao bound remains to date inaccessible.

\begin{figure*}[!t]
\centering
\includegraphics[width=0.9 \textwidth]{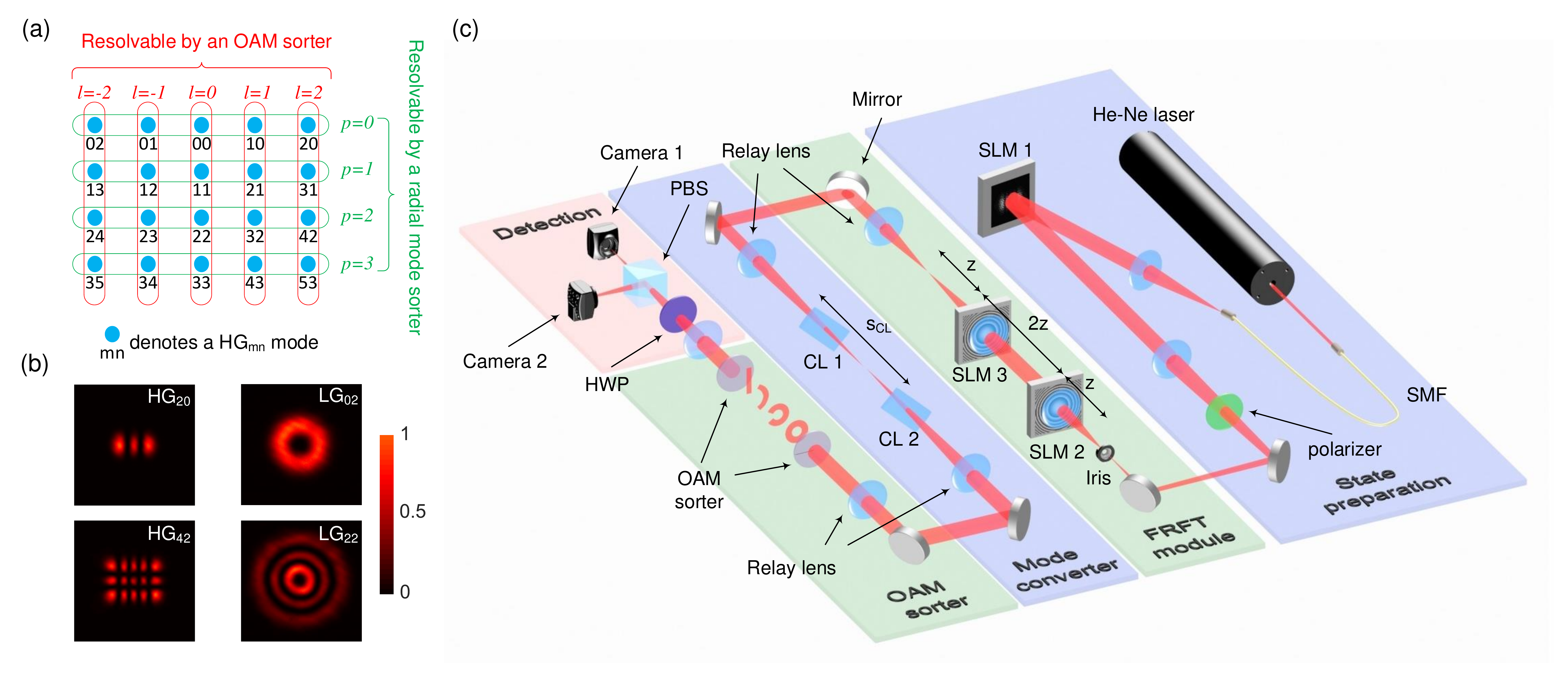}
\caption{(a) Conversion relation between HG modes and LG modes. Each dot represents a HG$_{mn}$ mode and the number under the dot denotes the mode indices $mn$. The HG modes in the same column are converted to LG modes of the same OAM index $\ell$, and the modes in the same row correspond to the LG modes of the same radial index $p$. (b) Experimental results of mode conversion from HG$_{20}$ to LG$_{02}$ and from HG$_{42}$ to LG$_{22}$. (c) Schematic of the HG mode sorter. A binary grating on SLM1 generates HG modes in the first diffraction order that are then analyzed by the rest of the setup. The FRFT module consists of two SLMs and two lens. Two cylindrical lenses (CLs) form the $\pi/2$ mode converter. The OAM sorter consists of an unwrapper and a phase corrector to perform Cartesian to log-polar transformation.}
\label{fig:setup1}
\end{figure*}

Despite the usefulness of HG modes in various areas, practical applications have been so far impeded by the difficulty in efficient detection of these modes. In contrast, the sorting of the Laguerre-Gaussian (LG) modes, the rotationally-symmetric counterparts of HG modes, have been experimentally demonstrated \cite{mirhosseini2013efficient, berkhout2010efficient,lavery2012refractive,leach2002measuring,PhysRevLett.119.263602}. The azimuthal structure of LG modes is found to be directly related to the orbital angular momentum (OAM) of photons \cite{allen1992orbital}, and it has been shown that a Cartesian to log-polar transformation can enable an OAM mode sorter to identify a large number of OAM modes \cite{mirhosseini2013efficient, berkhout2010efficient,lavery2012refractive}. Nonetheless, this method is not directly applicable to the HG modes. Several previous investigations on the sorting of HG modes are based on cascaded Mach-Zenhder interferometers \cite{ xue2001beam, linares2017interferometric}, and the practicality of such approaches greatly limits the number of HG modes that can be sorted. The multi-plane light conversion method has also been proposed to sort the spatial modes \cite{Fontaine2018Laguerre}. While this method is able to separate many modes simultaneously, the crosstalk between neighboring modes is relatively high and a large number of planes are usually needed to reduce the crosstalk. Therefore it remains highly desirable to build a robust sorter that can efficiently separate many HG modes with low crosstalk.

Here we describe a method for efficiently sorting HG modes. This method entails converting a HG mode to a unique LG mode and then sorting the LG mode with known methods. This sorter operates by sequentially applying a fractional Fourier transform (FRFT) module \cite{PhysRevLett.119.263602}, an astigmatic mode converter \cite{beijersbergen1993astigmatic}, and an OAM mode sorter. Our scheme takes advantage of a useful relation between HG modes and LG modes: the conversion between these two families of modes can be realized by an astigmatic mode converter \cite{beijersbergen1993astigmatic}. This mode converter can be implemented by two cylindrical lenses, and it has been shown that such a converter can transform $\text{HG}_{mn}$ to $\text{LG}_{p \ell}$ conditioned on $p=\min (m,n)$ and $\ell=m-n$, where $m$ and $n$ are the mode indices of a HG mode along $x$ and $y$ directions, and $p$ and $\ell$ are the radial and azimuthal indices of a LG mode. Therefore, when we cascade an astigmatic mode converter and an LG mode sorter, we can realize a HG mode sorter. In Fig.~\ref{fig:setup1} (a) the conversion relationship between HG modes and LG modes is visualized, and the experimental results of two HG-LG mode conversions are shown in Fig.~\ref{fig:setup1} (b). The HG modes in different columns in Fig.~\ref{fig:setup1} (a) can be converted to LG modes of different OAM indices, and therefore can be resolved by an OAM sorter. For example, we note that $\text{HG}_{m0}$ ($\text{HG}_{0n}$)  can be transformed to $\text{LG}_{0m}$ ($\text{LG}_{0,-n}$), and therefore one can readily utilize an OAM mode sorter to unambiguously sort HG modes with one zero index. This result leads to significant consequences, such as for super-resolution imaging in microscopy, as an efficient sorting of HG$_{00}$, HG$_{01}$, and HG$_{10}$ modes can resolve sub-diffraction objects and attain the Cram\'er-Rao bound \cite{tsang2016quantum}.

In order to determine both mode indices $m$ and $n$ simultaneously, however, the knowledge of $\ell=m-n$ is not sufficient. In other words, another index $p=\min (m,n)$ needs to be identified to completely characterize the LG mode converted from a HG mode. Here we resort to the FRFT as a secondary sorting mechanism, which has been recently used to sort the radial index of LG modes \cite{PhysRevLett.119.263602, yiyu2018quantum}. Since the HG modes are the eigenmodes of FRFT, they keep invariant under such a transform except for a mode-dependent Gouy phase. This transformation can be written as \cite{almeida1994fractional}
\begin{equation}\label{Eq:FGphase}
\mathcal{F}^{\alpha } [\text{HG}_{mn}(x_0,y_0) ] =\exp[{-i(m+n)  {\alpha}}] \cdot \text{HG}_{mn}(x,y),
\end{equation}
where $\mathcal{F}^{\alpha }$ denotes an FRFT of order $\alpha$, and HG$_{mn}(x,y)$ is the transverse field distribution at the beam waist plane which can be expressed as
\begin{equation}\label{Eq:HG}
\text{HG}_{mn}(x,y)=C_{mn} H_m \left( \frac{\sqrt{2}x}{w_0} \right) H_n \left( \frac{\sqrt{2}y}{w_0} \right) e^{-(x^2+y^2)/w_0^2},
\end{equation}
\begin{figure}[t]
\center
\includegraphics[width=0.5 \linewidth]{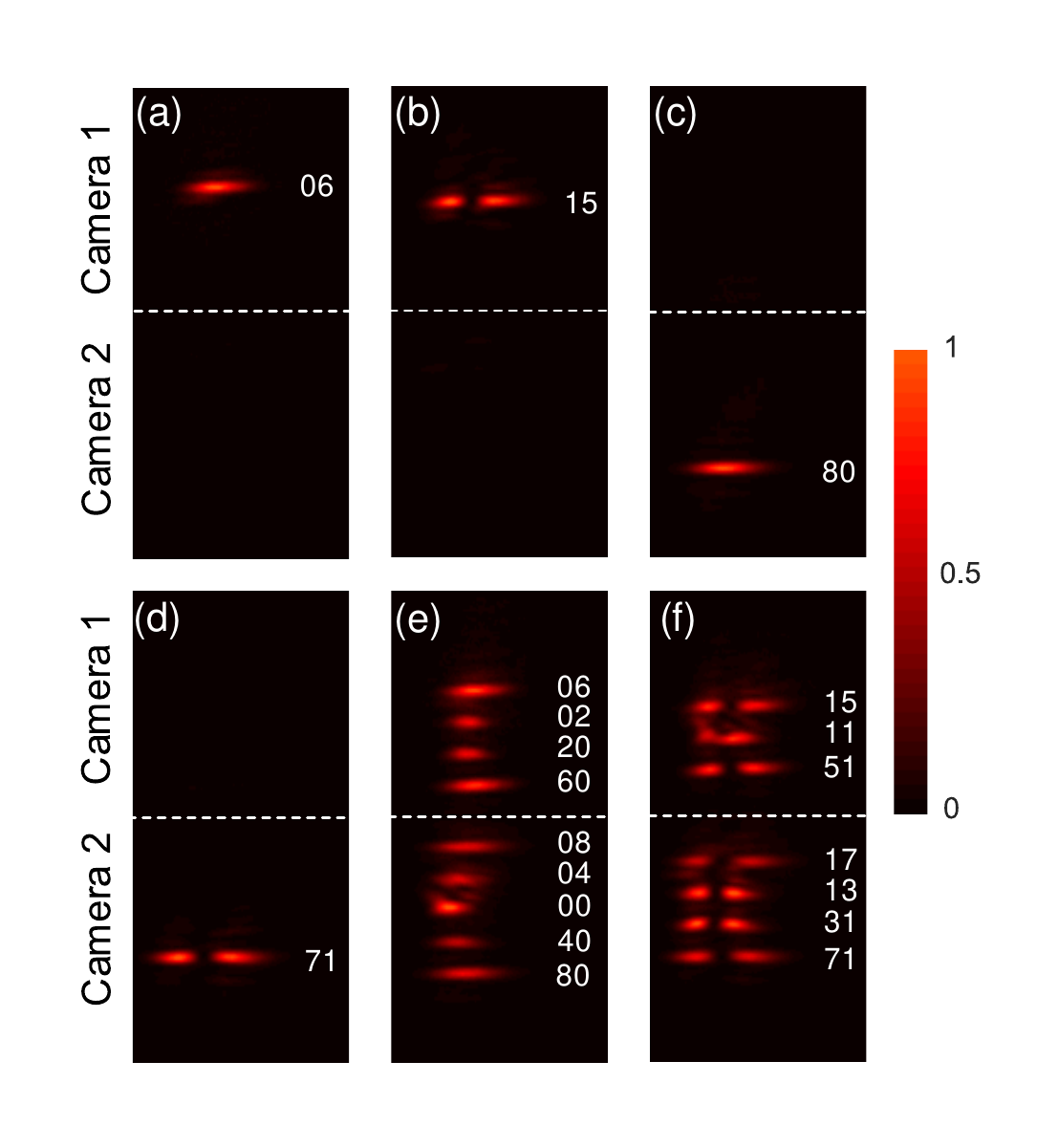}
\caption{Experimental results of the HG mode sorter. (a)-(d) Intensity distribution on two cameras when individual HG modes are injected. The mode indices $mn$ are labelled beside each sorted mode. (e) Combined intensity distribution on two cameras when incident modes include HG$_{0n}$ of $n=0,2,\cdots,8$ and HG$_{m0}$ of $m=0,2,\cdots,8$. (f) Combined intensity distribution at two output ports when incident modes include HG$_{1n}$ of $n=1,3,5,7$ and HG$_{m1}$ of $m=1,3,5,7$. }
\label{fig:sorterresult}
\end{figure}
where $C_{mn}$ is some normalization factor, $H_m(\cdot)$ and $H_n(\cdot)$ are the Hermite polynomials of order $m$ and $n$ respectively, and $w_0$ is the beam waist radius. One can notice that the Gouy phase $\exp[{-i(m+n)  {\alpha}}]$ contains information regarding the mode order $m+n$, and it has been demonstrated that an interferometer can be built to sort beams of different mode order $m+n$ to different output ports \cite{ xue2001beam, linares2017interferometric}. In our experiment, we implement a common-path interferometer by using polarization-dependent SLMs to realize an inherently stable sorter \cite{yiyu2018quantum}. The horizontal and vertical polarizations are employed as two arms of a Mach-Zenhder interferometer, and two polarization-sensitive SLMs are used to perform FRFT of different orders to the two polarizations respectively. If the FRFT order difference is $\Delta \alpha$, then a mode-dependent phase of $\Delta \phi=-(m+n)\cdot \Delta \alpha$ is introduced between the two polarizations at the output of the FRFT module. By measuring the polarization state of the output photons, one can determine the value of $m+n$. Due to the bounded two-dimensional Hilbert space of polarization, our FRFT module and the following OAM mode sorter can only resolve the HG modes located at the top two rows in Fig.~\ref{fig:setup1}(a). However, the sorting capability can be readily extended by cascading more FRFT modules as will be discussed later.

\begin{figure}[t]
\center
\includegraphics[width= 0.6 \linewidth]{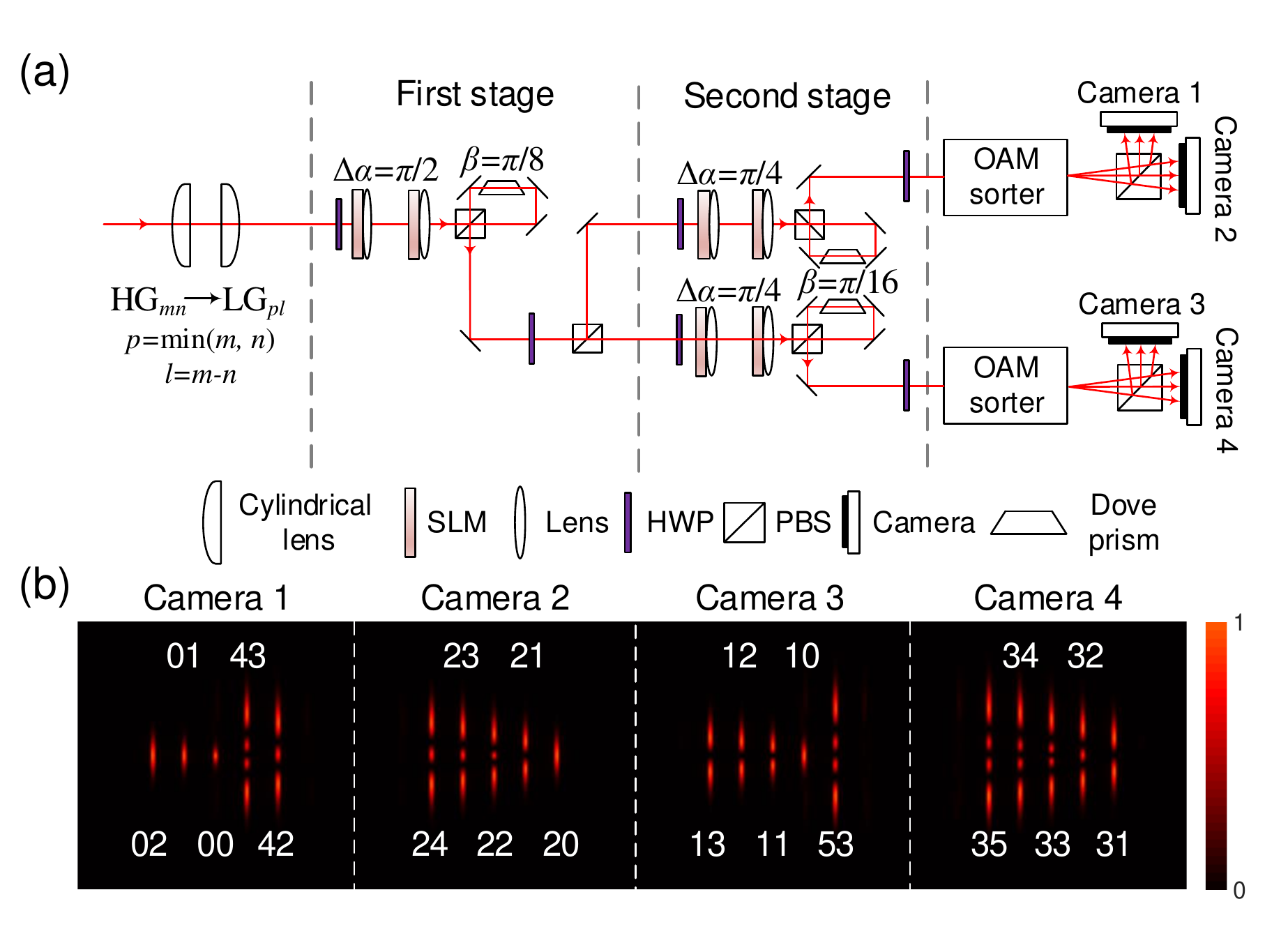}
\caption{(a) Conceptual schematic of HG mode sorter of an extended dimension. A HG$_{mn}$ is first transformed to LG$_{p\ell}$, and then FRFT modules cascaded by a Dove prism inside a Sagnac interferometer are applied to sort the LG mode. Relay lenses to prevent spatial modes from diffraction are omitted for simplicity. (b) Simulated sorting result for entire 20 HG modes which are listed in Fig.~\ref{fig:setup1}(a). Each mode is normalized to have the same maximum intensity. It can be noticed that each mode is mapped to distinct position on cameras unambiguously. The mode index $mn$ is labeled around the corresponding sorted mode.}
\label{fig:scalable}
\end{figure}

The experimental schematic of our HG mode sorter is shown in Fig.~\ref{fig:setup1} (c). A 633~nm He-Ne laser is spatially filtered by a single mode fiber (SMF) and then collimated to illuminate SLM~1. A computer-generated hologram is imprinted on SLM 1 to generate HG modes in the first diffraction order \cite{mirhosseini2013rapid}. A polarizer sets the light to be diagonally polarized. The beam waist radius of HG modes is $462.3$ $\mu$m. A quadratic phase equivalent to that of a 0.62 m lens is imprinted on SLM~2 and SLM~3. Each SLM is attached with a lens of focal length 1.5~m, and in the experiment we realize this by relaying each lens to the corresponding SLM via a 4-$f$ system respectively. Both SLMs are only effective to horizontal polarization and do not modulate the vertical polarization. The free-space propagation distance between SLM2 and SLM3 is chosen to be $2z$, where $z=0.44$~m. It can be verified that SLM~2, SLM~3, and two spherical lenses implements an FRFT of $\alpha=\pi/2$ to vertical polarization \cite{yiyu2018quantum}. For horizontal polarization, these elements act as a 4-$f$ system of $\alpha=\pi$. According to Eq.~(\ref{Eq:FGphase}), the diagonally polarized input HG modes become diagonally (anti-diagonally) polarized at the output when the value $(m+n)/2$ is even (odd). After this module, two relay lenses resize the beam waist radius to match an astigmatic mode converter constituted by a pair of 45-degree oriented cylindrical lenses. The focal length of each cylindrical lens $f_{CL}$,  the beam waist radius $w_0$, and the separation between two cylindrical lenses $s_{CL}$ are related by the following equations \cite{beijersbergen1993astigmatic}:
\begin{equation}
w_0 = \sqrt{\frac{(1+1/\sqrt{2})\lambda f_{CL}}{\pi}}, \hspace{1cm} s_{CL}=\sqrt{2}f_{CL},
\label{eq:CLlens}
\end{equation}
where $\lambda=633$ nm is the laser wavelength. In our experiment we use $f_{CL}=10$ cm and $s_{CL}=14.1$ cm, and the consequent beam waist radius is $w_0=185.5 \text{ }\mu$m. The transformed HG modes, which have become LG modes at this point, are sent to a polarization-independent OAM mode sorter \cite{lavery2012refractive}. A half wave-plate (HWP) and a polarizing beamsplitter (PBS) directs photons to different output ports according to their orthogonal polarization assigned by the FRFT module.

The experimental results of our HG mode sorter are shown in Fig.~\ref{fig:sorterresult}. In Fig.~\ref{fig:sorterresult}(a)-(d) we present the images on two cameras when HG modes are injected into the sorter individually. In Fig.~\ref{fig:sorterresult}(e) we combine the sorting result of HG$_{0n}$ with $n=0,2,\cdots,8$ as well as HG$_{m0}$ with $m=0,2,\cdots,8$. The lowest index of these modes is 0, which is represented by the first row in Fig.~\ref{fig:setup1}(a). We choose the index spacing to be 2 to reduce the overlap between neighboring modes for the purpose of better visualization. We emphasize that there is no fundamental restriction on the index spacing provided that the OAM mode sorter can have a sufficient mode resolution \cite{mirhosseini2013efficient}. One can notice that the HG modes for which $(m+n)/2$ is even (odd) are routed to camera 1 (camera 2) as a result of the FRFT module. As mentioned earlier, one FRFT module and one OAM mode sorter can resolve the modes in the top two rows in Fig.~\ref{fig:setup1}(a), and the experimental evidence is presented in Fig.~\ref{fig:sorterresult}(f). We generate HG$_{1n}$ with $n=1,3,5,7$ and  HG$_{m1}$ with $m=1,3,5,7$, and combine the corresponding images from two cameras. The lowest index of these modes is 1, which are represented by the second row in Fig.~\ref{fig:setup1}(a). Notably, one can see that  these two sets of modes occupy different positions on cameras and in principle can be fully identified by a high-resolution OAM sorter \cite{mirhosseini2013efficient}. However, we also note that our setup cannot separate HG modes whose mode order $m+n$ is odd. For example, one can verify that HG$_{10}$ and HG$_{21}$ cannot be separated by either FRFT module or the subsequent OAM mode sorter. This problem can be addressed by cascading a Dove prism and a Sagnac interferometer to the FRFT module \cite{2018Realization}. Here we provide a conceptual design and demonstrate how to extend the dimension of a HG mode sorter. As shown in Fig.~\ref{fig:scalable}(a), the input linearly polarized HG$_{mn}$ is converted to LG$_{p\ell}$ and sent to the subsequent LG mode sorter. {Compared to Fig.~\ref{fig:setup1}, we place the astigmatic mode converter before the FRFT module due to the consideration of simplicity. The input HG$_{mn}$ mode and the converted LG$_{p\ell}$ mode is related by $m+n=2p+|\ell|$, and the phase induced by FRFT is a function of $m+n$ for HG modes and $2p+|\ell|$ for LG modes. Therefore swapping the mode converter and the FRFT module does not influence the sorting mechanism.} Each FRFT module is followed by a Dove prism inside a Sagnac interferometer. The FRFT module induces a phase difference $\Delta \psi_1=-(m+n)\cdot \Delta \alpha$ to horizontal and vertical polarizations \cite{PhysRevLett.119.263602}, and the Dove prism can rotate horizontally and vertically polarized LG modes by an angle of $2\beta$ and $-2\beta$ respectively, where $\beta$ is the orientation angle of the Dove prism. This polarization-dependent rotation leads to a phase difference of $\Delta \psi_2=-4\beta(m-n)$ between horizontal and vertical polarizations \cite{leach2002measuring}, where $m-n=\ell$ represents the OAM index of the converted LG modes. Therefore the total phase difference is $\Delta \psi = -[(\Delta \alpha+4\beta)m+(\Delta \alpha-4\beta)n]$ and one can realize an unambiguous sorting with appropriately selected $\Delta \alpha$ and $\beta$. These parameters used in each stage are provided in Fig.~\ref{fig:scalable}(a). It can be verified that all HG modes will now be guided towards a unique, mode-dependent output port \cite{2018Realization}, and the output port of each mode can be predicted by Eq.~(\ref{Eq:FGphase}). We numerically simulate the output image of this extended sorter, and the combined simulation result for 20 HG modes is presented in Fig.~\ref{fig:scalable}(b). These 20 HG modes are the modes listed in Fig.~\ref{fig:setup1}(a) with a mode index spacing of 1. In the simulation we use a beam-copying grating in an OAM sorter, which can create 7 copies of a beam to improve the mode resolution \cite{mirhosseini2013efficient}. It can be noticed that these 20 HG modes can be well separated to distinct positions unambiguously with negligible crosstalk. {As demonstrated in \cite{mirhosseini2013efficient}, with a beam-copying grating the separation efficiency can be improved to theoretically 97\%, and experimentally achieved efficiency can be larger than 92\%. We also emphasize that by removing the astigmatic LG-HG mode converter, this sorter becomes a LG mode sorter and can unambiguously separate LG modes of $p \in \{0,1,2,3\}$ and arbitrary $\ell$.}

{In our experiment the loss mainly comes from the SLMs due to the limited light utilization efficiency, which might impede the scaling shown in Fig.~\ref{fig:scalable}. However, we note that the SLMs can be readily replaced by other low-loss devices, such as the commercially available polarization directed flat lenses \cite{2018Realization}. In addition, since all devices employed in our scheme are essentially phase-only elements, the loss can in principle be reduced to zero if appropriate anti-reflection coating is applied on all elements. However, given that the polarization has been used in our scheme to realize a robust FRFT module, our sorter cannot work directly for an arbitrary polarization. This limitation can be lifted e.g., by using a polarization-independent FRFT module \cite{PhysRevLett.119.263602} or by inserting a PBS to separate polarizations before the sorter. Moreover, the sorting scheme presented here can in principle be used to HG modes of different wavelengths. As can be seen in Eq.~(\ref{eq:CLlens}), the parameters $f_{CL}$ and $s_{CL}$ remains the same as long as $w_0^2/\lambda$ keeps constant, which is also true for the parameters of the FRFT module (see Eq. (3) in \cite{PhysRevLett.119.263602}). We also note that the material dispersion is the only factor that limits the spectral bandwidth of OAM mode sorter \cite{lavery2012refractive}. Therefore, our sorter in principle can have a relatively broad bandwidth provided that $w_0^2/\lambda$ is constant and material dispersion is small.}

In conclusion, we have proposed and experimentally realized a scalable scheme that can efficiently sort a large number of HG modes. Our scheme is based on an astigmatic mode converter to transform HG modes to LG modes and takes advantage of a LG mode sorter to realize a mode sorter for a large number of HG modes. Further increasing the dimension is straightforward and a conceptual schematic has been presented and numerically simulated. Taking into account the broad use of HG modes, we expect that our demonstration can facilitate fundamental studies of beam analysis in free space and graded-index multi-mode fiber, and can enhance a variety of applications such as super-resolution imaging, optical communications, and quantum key distribution.

\begin{acknowledgments}
U.S. Office of Naval Research (grant No. N000141712443 and N000141512635). R. W. B. acknowledges support from Canada Excellence Research Chairs Program and Natural Science and Engineering Research Council of Canada.
\end{acknowledgments}

%


\end{document}